\documentclass[prd,aps,floats,twocolumn,twoside,preprintnumbers,
superscriptaddress,floatfix,nofootinbib,showpacs]{revtex4}
\usepackage{graphicx,color,amsmath,amssymb,hyperref,epsfig,tensor}
\usepackage{pstricks,rotating}
\usepackage{slashed}

\def\eq#1\en{\begin{equation} #1 \end{equation}}

\def\eqa#1\ena{\begin{eqnarray} #1 \end{eqnarray}}

\DeclareMathOperator{\Tr}{Tr}

\makeatletter
\newcommand{\fmslash}[2][0mu]{%
  \mathchoice
    {\fmsl@sh\displaystyle{#1}{#2}}%
    {\fmsl@sh\textstyle{#1}{#2}}%
    {\fmsl@sh\scriptstyle{#1}{#2}}%
    {\fmsl@sh\scriptscriptstyle{#1}{#2}}}
\newcommand{\fmsl@sh}[3]{%
  \m@th\ooalign{$\hfil#1\mkern#2/\hfil$\crcr$#1#3$}}

\newcommand{\tr}{\hbox{tr}}

\begin{document}


\title{Forbidden and invisible Z boson decays in a covariant\\ $\theta$-exact noncommutative standard model}

\author{R. Horvat}
\affiliation{Institute Rudjer Bo\v{s}kovi\'{c}, Bijeni\v{c}ka 54, 10000 Zagreb, Croatia}
\author{A. Ilakovac}
\affiliation{Faculty of Science, University of Zagreb, Bijeni\v{c}ka 32, 10000 Zagreb, Croatia}
\author{D. Kekez}
\affiliation{Institute Rudjer Bo\v{s}kovi\'{c}, Bijeni\v{c}ka 54, 10000 Zagreb, Croatia}
\author{J. Trampeti\'{c}}
\affiliation{Institute Rudjer Bo\v{s}kovi\'{c}, Bijeni\v{c}ka 54, 10000 Zagreb, Croatia}
\affiliation{Max-Planck-Institut f\"ur Physik,
	(Werner-Heisenberg-Institut),
  	 F\"ohringer Ring 6, D-80805 M\"unchen, Germany}
\author{J. You}
\affiliation{Institute Rudjer Bo\v{s}kovi\'{c}, Bijeni\v{c}ka 54, 10000 Zagreb, Croatia}
\affiliation{Mathematisches Institut G\"ottingen, Bunsenstr. 3-5,
	37073 G\"ottingen, Germany}


\begin{abstract}
The triple neutral gauge boson and direct $\rm{U(1)_Y}$-neutrino interactions, being absent in ordinary field theory, can arise quite naturally in noncommutative gauge field theories. Using non-perturbative methods and a Seiberg-Witten map based covariant approach to noncommutative gauge theory, we have found $\theta$-exact expressions for both interactions, thereby eliminating previous restrictions to low-energy phenomena. In particular we obtain for the first time the covariant, $\theta$-exact, triple neutral gauge boson interactions within the noncommutative Standard Model gauge sector including an additional gauge-field deformation freedom. Finally we discuss implications for $Z \to \gamma\gamma$ and $Z \to \bar\nu\nu$ decays, and show that our results behave quite reasonably throughout all interaction energy scales.
\end{abstract}



\maketitle


\section{Introduction}

After the string theory had indicated that noncommutative (NC) gauge field theory (GFT) could be one of its low-energy effective-theory realisation \cite{Seiberg:1999vs}, the studies on noncommutative particle phenomenology started their development \cite{Hinchliffe:2002km,Trampetic:2008bk}. The aim of the present paper is to find possible experimental signatures and predict/estimate bounds on the NC spacetime from collider physics experimental data, in particular from the Standard Model (SM) forbidden, earlier introduced and very well known $Z \to \gamma\gamma$ decay mode as well as  the invisible $Z \to \bar\nu\nu$ decay. The NC contributions are for the latter decay are analyzed and discussed first time in this article.

Nice progress has been obtained in the Seiberg-Witten (SW) maps
and enveloping algebra based models where one could deform commutative
gauge theories with arbitrary gauge group and representation~
\cite{Madore:2000en,Jurco:2000ja,Jurco:2000fb,Jurco:2001my,Jurco:2001rq,
Jackiw:2001jb,Calmet:2001na,Horvat:2011qn}. Possible constraints on the $\rm U_\star(1)$ 
charges \cite{Chaichian:2009uw} are also rescinded in this approach.
The noncommutative extensions of particle physics models like the covariant noncommutative SM (NCSM) and the NC GUT models \cite{Calmet:2001na,Horvat:2011qn,Behr:2002wx, Deshpande:2001mu,Duplancic:2003hg,Aschieri:2002mc,Melic:2005fm,Martin:2013gma,Martin:2013lba} were constructed.
These allow a minimal (no new particle content) deformation with
the sacrifice that interactions include infinitely many terms defined
through recursion over (and in practice the cut-off at a certain order of) the NC parameter $\theta^{\mu\nu}$. There were also other studies with NC modifications of the particle-physics theory \cite{Bichl:2001cq,Ohl:2004tn,Melic:2005su,Alboteanu:2006hh,Buric:2007qx}.

Studies of divergences have been performed since the early stage of
the NC QFT development \cite{Filk:1996dm}.  Later on trouble-free one-loop quantum corrections to noncommutative scalar $\phi^4$ theories \cite{Grosse:2004yu,Magnen:2008pd,Meljanac:2011cs} and the NC QED \cite{Vilar:2009er} have been obtained.
Also the SW map expanded NCSM \cite{Calmet:2001na,Behr:2002wx,Duplancic:2003hg,Melic:2005fm} at first order in $\theta$, albeit breaking the Lorentz symmetry is found to be anomaly free \cite{Martin:2002nr}, having also  well-behaved one-loop quantum corrections \cite{Bichl:2001cq,Buric:2006wm,Latas:2007eu,Buric:2007ix, Martin:2009vg,Martin:2009sg,Tamarit:2009iy,Buric:2010wd}.

At $\theta$-order there are two important interactions that are suppressed
and/or forbidden in the SM, the triple neutral gauge boson interaction
\cite{Behr:2002wx,Duplancic:2003hg,Melic:2005fm}, and the tree level coupling of neutrinos to photons \cite{Schupp:2002up, Minkowski:2003jg}, respectively. Here the expansion and cut-off in powers of the NC parameters $\theta^{\mu\nu}$ corresponds to an expansion in momenta, restricting the range of validity to energies well below the NC scale $\Lambda_{\rm NC}$. Usually, this is no problem for experimental predictions because the lower bound on the NC parameters $\theta^{\mu\nu}=c^{\mu\nu}/\Lambda_{\rm NC}^2$ (the coefficients $c^{\mu\nu}$ being of order one) runs higher than typical momenta involved in a particular process. However, there are exotic processes in the early universe as well as those involving ultra high energy cosmic rays \cite{Horvat:2009cm,Horvat:2010sr,Horvat:2011iv,Horvat:2011wh} in which the typical energy involved is higher than the current experimental bound on the NC scale $\Lambda_{\rm NC}$. Thus, the previous $\theta$-cut-off approximate results are inapplicable. To cure the cut-off approximation, we are using $\theta$-exact expressions, inspired by the exact formulas for the SW map \cite{Jurco:2001my,Mehen:2000vs,Liu:2000mja,Okawa:2001mv,Martin:2012aw}, and expand in powers of gauge fields, as we did in \cite{Horvat:2011iv}. In $\theta$-exact models we have studied the UV/IR mixing \cite{Schupp:2008fs,Horvat:2011bs}, the neutrino propagation \cite{Horvat:2011qg} and also some NC photon-neutrino phenomenology \cite{Horvat:2009cm,Horvat:2010sr,Horvat:2011iv,Horvat:2011wh}, respectively.

Due to the presence of the UV/IR mixing the $\theta$-exact model is generally regarded as not perturbatively renormalizable,  thus the relation of quantum corrections to observations is not entirely clear~\cite{Horvat:2010km}. However, recently we observed that certain possibility of reducing the severity of divergences using constraints on both the deformation freedom and noncommutative parameter $\theta$ \cite{Horvat:2011bs,Horvat:2011qg,Horvat:2013rga} could occur. In a noncommutative $\rm U(1)$ model we found that via the combination of deformation freedom and appropriate choices of the noncommutative parameter $\theta$, one could remove all divergences from both the one-loop neutrino (neutral fermion) and photon self-energy. In particular there exists a particular combination where the photon self-interaction contribution to photon self-energy reduces to a single finite term~\cite{Horvat:2013rga}. A deformation freedom was also shown recently to be capable of suppressing the divergent structures at one-loop and first order in the noncommutative parameter $\theta$~\cite{Buric:2006wm,Latas:2007eu,Buric:2007ix, Martin:2009vg,Martin:2009sg,Tamarit:2009iy,Buric:2010wd}.\footnote{Photon self-energy can be controlled at any order of $\theta$-expansion when such parameters are allowed to be renormalized~\cite{Bichl:2001cq}. It may also worth notice that in a $\theta$-exact treatment the momentum power series sum into phase shifts, therefore do not increase the superficial UV divergent order.} Assuming that controls over divergences in noncommutative phenomenological models could be made available, either in the $\theta$-expanding or in the $\theta$-exact approach, tree level amplitudes will always serve as the leading order contribution to the relevant particle physics process in such models given perturbation theory is trustable.

In the particular context of the UV/IR mixing problem it is also very important to mention a complementary  approach \cite{hep-th/0606248,Abel:2006wj} where NC gauge theories are realized as effective QFTs, underlain by some  more fundamental theory such as string theory. It was claimed that for a large class of more general QFTs above the UV cutoff the phenomenological effects of the UV completion can be quite successfully modeled by a threshold value of the UV cutoff. So, in the presence of a finite UV cutoff {\em neither} sort of divergence will ever appear since the problematic phase factors effectively transform the highest energy scale (the UV cutoff) into the lowest one (the IR cutoff). What is more, not only the full scope of noncommutativity is experienced only in the range delimited by the two cutoffs, but for the scale of NC high enough, the whole standard model can be placed below the IR cutoff. In such a way the UV/IR mixing problem becomes hugely less pressing, making, at the same time, a study of the theory at the quantum level much more reliable.

To study triple neutral gauge boson couplings,
in this report we construct the $\theta$-exact pure gauge-sector
action of the $\rm SU(2)\times U(1)$ gauge group, while for
the NC Z-boson-neutrino and photon-neutrino interactions we are using
the actions from \cite{Horvat:2011qn}. We compute only tree level processes
and therefore treat the NC model as an effective theory only. The decay of $Z$ boson into two photons has been used to predict possible experimental signatures of the NCSM \cite{Behr:2002wx,Duplancic:2003hg,Melic:2005fm,Buric:2007qx}. Since by Bose-symmetry and rotational invariance arguments \cite{LandauYoung} any vector particle cannot
decay into two massless vector particles, this forbidden decay has very little background from the standard model. Fixing $\theta$ spontaneously breaks C, P, and/or CP discrete symmetries \cite{Aschieri:2002mc}. See also general discussions on the C,P,T, and CP properties of the noncommutative interactions  in \cite{SheikhJabbari:2000vi}, and in the case of our model a discussion  given in \cite{Melic:2005hb,Tamarit:2008vy}.  A breaking of C symmetry occurs in the $Z\to \gamma\gamma$ process.  One common approximation in those existing works is that only the vertices linear in the NC parameter $\theta$ were used. 

In this work we extend the NCSM gauge sector actions for first time to all orders in  $\theta$. Next we discuss the decay widths $\Gamma(Z\to \gamma\gamma)$  and $\Gamma(Z\to \nu\nu)$ as functions of the NC scale $\Lambda_{\rm NC}$ for space/light-like noncommutativity which are allowed by the unitarity conditions \cite{Gomis:2000zz,Aharony:2000gz}.

\section{NCSM gauge sector in a $\theta$-exact model of the NCSM}

As usual we consider the star$(\star)$-product formalism for quantum field theory on
the deformed Moyal space with a constant NC parameter $\theta$. We start with a $\theta$-exact hybrid SW-map expansion of the noncommutative gauge field $\hat V_\mu$ in the terms of component commutative fields,
\begin{eqnarray}
\hat V_\mu
&=&
V_\mu^aT^a-\frac{1}{8}\theta^{\rho\tau}
\bigg[\Big\{V_\rho^a\stackrel{\star_2}{,}(\partial_\tau
V_\mu+F_{\tau\mu})^b\Big\}\big\{T^a,T^b\big\}
\nonumber\\
&&+\:
\Big\{V_\rho^a\stackrel{\star_{2'}}{,}(\partial_\tau
V_\mu+F_{\tau\mu})^b\Big\}
\big[T^a,T^b\big]
\bigg].
\label{compfield}
\end{eqnarray}
Besides the star($\star$)-commutator term, starting at first order of $\theta$,
one gets a $\star$-anti\-commutator term starting at second order in $\theta$.
The generalized star products: $\star_2$ and $\star_{2'}$ \cite{Mehen:2000vs,Schupp:2008fs,Horvat:2011bs,Horvat:2011qn}, have the following definitions and properties:
\begin{eqnarray}
\phi(x)\star_2 \psi(x)&=&\psi(x)\star_2 \phi(x)
\nonumber\\
&=&\frac{\sin\frac{\partial_1\theta
\partial_2}{2}}{\frac{\partial_1\theta
\partial_2}{2}}\phi(x_1)\psi(x_2)\bigg|_{x_1=x_2=x},
\nonumber\\
\big[\phi\stackrel{\star}{,}\psi\big]
&=&i\theta^{\rho\tau}\partial_\rho\phi\star_2\partial_\tau\psi,
\label{star2}
\end{eqnarray}
\begin{eqnarray}
\phi(x)\star_{2'}\psi(x)&=&-\psi(x)\star_{2'} \phi(x)
\nonumber\\
&=&\frac{\cos\frac{\partial_1\theta\partial_2}{2}-1}
{\frac{\partial_1\theta\partial_2}{2}}\phi(x_1)\psi(x_2)\bigg|_{x_1=x_2=x},
\nonumber\\
\big\{\phi \stackrel{\star}{,} \psi\big\}-\big\{\phi,\psi\big\}&=&\theta^{\rho\tau}\partial_\rho\phi\star_{2'}\partial_\tau\psi,
\label{star2'}
\end{eqnarray}
where $\star_2$ is symmetric and $\star_{2'}$ antisymmetric in its arguments. Both star products are non\-associative.

The noncommutative action is defined in the usual way \cite{Buric:2006wm,Schupp:2002up,Minkowski:2003jg,Schupp:2008fs}
\begin{equation}
S=\int-\frac{1}{2}\Tr \hat F^{\mu\nu}\star \hat F_{\mu\nu}+i \bar{\hat \Psi}\star\fmslash{D}\hat \Psi\,,
\label{Sg1}
\end{equation}
with definitions of the noncommutative covariant derivative and field strength resembling the corresponding expressions of non-abelian Yang-Mills theory:
\begin{eqnarray}
D_\mu\hat \Psi&=&\partial_\mu\hat \Psi-i[\hat V_\mu\stackrel{\star}{,}\hat \Psi],
\nonumber\\
\hat F_{\mu\nu}&=&\partial_\mu \hat V_\nu-\partial_\nu
\hat V_\mu-i[\hat V_\mu\stackrel{\star}{,}\hat V_\nu].
\label{DF}
\end{eqnarray}
Noncommutative fields ($\hat V_\mu,\hat \Psi$) in above action are images of the corresponding commutative fields $V_\mu$ and $\Psi$ under hybrid SW-map. 

Starting with equations (\ref{compfield}) and (\ref{DF}), from the action (\ref{Sg1}) we obtain the following  gauge-action-part written in terms of the commutative component gauge fields:
\begin{eqnarray}
S_{g}&=&\frac{-1}{2}
\int\sum\limits_{a,b}\tr(T_aT_b)F^a_{\mu\nu}F^{b\mu\nu}
-\int\sum\limits_{a,b,c} \Bigg\{\tr T_a\{T_b,T_c\}
\nonumber\\
&\cdot& F^{a\mu\nu}\Bigg[\theta^{\rho\tau}\partial_\mu\bigg(V^b_\rho\star_2(\partial_\tau V^c_\nu
+F^c_{\tau\nu})\bigg)
+i[V^b_\mu\stackrel{\star}{,}V^c_\nu]\Bigg]
\nonumber\\
&+&\tr T_a[T_b,T_c]F^{a\mu\nu}
\Bigg[i\bigg(\{V^b_\mu\stackrel{\star}{,}V^c_\nu\}-\{V^b_\mu,V^c_\nu\}\bigg)
\nonumber\\
&-&
\theta^{\rho\tau}V^b_\rho\star_{2'}(\partial_\tau
V^c_\nu+F^c_{\tau\nu})\Bigg]\Bigg\}.
\label{Sg2}
\end{eqnarray}
The two traces $\tr T_a[T_b,T_c]$ and $\tr T_a\{T_b,T_c\}$
are both well known in the representation theory of Lie algebras.
The first one is proportional to the structure
constant $f_{abc}$, with the quadratic Casimir as its coefficient,
$
\tr T_a[T_b,T_c]=i\tr T_a f_{dbc}T_d=iA f_{abc}.
$
Thus the different gauge sectors do not mix in this part of the action.

The second  trace is slightly more
complicated as it is connected with details of representation of the
gauge group (we denote it as $B_{abc}$). Also, using the generalized star product $\star_2$, we can rewrite the star-commutator
as $i[V^b_\mu\stackrel{\star}{,}V^c_\nu]=-\theta^{\rho\tau}\partial_\rho
V^b_\mu\star_2 \partial_\tau V^c_\nu$, and obtain:
\begin{eqnarray}
S_{g}
&=&
\sum\limits_a A\int F^a_{\mu\nu}F^{a\mu\nu}
\nonumber\\
&-&
\sum\limits_{a,b,c}B_{abc}\int \theta^{\rho\tau}F^{a\mu\nu}
\Bigg[\partial_\mu\bigg(V^b_\rho\star_2(\partial_\tau V^c_\nu+F^c_{\tau\nu})\bigg)
\nonumber\\
&-&
\partial_\rho
V^b_\mu\star_2 \partial_\tau V^c_\nu
\Bigg]+...=S_{\rm gauge}+....
\label{Sg3}
\end{eqnarray}

After a series of partial integrations and by noting that $B_{abc}$ is totally symmetric under the permutations of $a$, $b$ and $c$, one arrives at
the action
\begin{eqnarray}
S_{\rm gauge}&=& A\int F^a_{\mu\nu}F^{a\mu\nu}
\label{Sg5}\\
-B_{abc}\hspace{-.7cm}&&\int \theta^{\rho\tau}F^{a\mu\nu}
\left(\frac{1}{4}F_{\rho\tau}^b\star_2F^c_{\mu\nu}-F^b_{\mu
\rho}\star_2 F^c_{\nu\tau}\right)+....
\nonumber
\end{eqnarray}
Here we see that the three gauge boson mixing coupling terms is controlled by the cubic Casimir $B_{abc}$ and contain only the generalized star-product $\star_2$. When $\star_2$ is reduced to unity by a $\theta$-expansion, this formula recovers the prior first order result in~\cite{Behr:2002wx}. $B_{abc}$ is in general group and representation dependent. It reflects a few common properties:\\
$\bullet$ $B_{abc}$ takes on the opposite sign for a representation and its complex conjugate;\\
$\bullet$ $B_{abc}$ vanishes for the adjoint representation of any Lie group;\\
$\bullet$ $B_{abc}$ vanishes for any simple Lie group except ${\rm SU(N}\ge 3)$.\\

For this reasons, $\rm SO(10)$ and $\rm E_6$ GUT models have
no additional noncommutative triple gauge boson
couplings which are forbidden in the standard model
(this attribute was qualified as the ``uniqueness'' of
the noncommutative GUT model in \cite{Aschieri:2002mc}).
Furthermore when the trace in the standard model is computed
using generators descended from non-$\rm SU(N\ge 3)$ simple gauge groups, for example from $\rm SO(10)$ or $\rm E_6$ GUTs, there is no such coupling either. The standard model and $\rm SU(5)$ GUT were studied before \cite{Deshpande:2001mu,Behr:2002wx,Aschieri:2002mc,Melic:2005fm} as both could accommodate the neutral boson mixing
couplings $\gamma\gamma\gamma$, $Z\gamma\gamma$,
$ZZ\gamma$, $\gamma GG$, $ZGG$, and $ZZZ$.
Since in $\rm SU(5)$ GUT all heavy gauge bosons are charged,
the standard model neutral boson coupling has covered all the cases.
To introduce possible extra neutral gauge bosons, one may consider models with left-right symmetry like the Pati-Salam
$\rm SU(4)\times SU(2)\times SU(2)$ \cite{Pati:1974yy}
or trinification $\rm SU(3)^3\times Z_3$ \cite{trinification}.
In the Pati-Salam model, the vanishing of $B_{abc}$ for $\rm SU(2)$
forces all mixing to arise from the $\rm SU(4)$ sector,
that is,  $YYY$, $YGG$ (and $GGG$) couplings only,
thus no mixing will include heavy gauge bosons.
The $\rm SU(4)$ sector has either $\bf 4$ or $\bf {\bar 4}$ representation.
Therefore, up to normalization the following terms
\begin{eqnarray}
\tr Y\{Y,Y\}&=&-\frac{8}{9},\;\;\;\tr Y \{G^a,G^b\}=\delta^{ab},\,\;
\nonumber\\
\tr G^a\{G^b,G^c\}&=&d^{abc}_{(3)},
\label{YYY}
\end{eqnarray}
are the only non-vanishing neutral-boson coupling components.
The trinification $\rm SU(3)^3\times Z_3$ seems more promising since
the left and right symmetry group are both $\rm SU(3)$.
However, in this model, all matter multiplets are of $\rm Z_3$
symmetric $(3,\bar 3,1)\oplus(\bar 3,1,3)\oplus(1,3,\bar 3)$ type.
Thus,  when $\rm Z_3$ is maintained, all mixing couplings cancel between $\bf 3$ and $\bf {\bar 3}$.\footnote{An analogues conclusion was drawn from
the analysis of the NCSM gauge sector at first order in $\theta$
\cite{Buric:2006wm}. Also note that to exploit the possibility of $Z'\gamma\gamma$ coupling one could probe, for example, the left-right symmetric
electroweak model $\rm SU(3)_L\times SU(3)_R\times U(1)_X$ \cite{Dias:2010vt}.}

In order to obtain the NCSM triple neutral gauge boson (TGB) interaction terms, in accordance with \cite{Calmet:2001na,Behr:2002wx,Deshpande:2001mu,Aschieri:2002mc,Duplancic:2003hg,Melic:2005fm}, we return to the standard model gauge group, $\rm G_{SM}=U(1)_Y \times SU(2)_L \times SU(3)_C$,  write the SM gauge potential as
\begin{eqnarray}
V^{\mu}=g_Y A^{\mu}Y + g_L\sum^3_{a=1}B_a^{\mu}T^a_L+g_C\sum^8_{b=1}G_b^{\mu}T^b_C\,,
\label{V}
\end{eqnarray}
and choose to sum in the action over all particle representations of the standard model. There are five multiplets of fermions and one Higgs multiplet for each generation, so we assign six arbitrary weights $\alpha_i,i=1...6$, to each of them. Considering triple-gauge boson couplings we find the following non-vanishing elements to be
\begin{eqnarray}
&\tr Y\{T^i_L,T^j_L\}=\tr T^i_L\{Y,T^j_L\}=\tr T^i_L\{T^j_L,Y\},
\nonumber\\&
\tr Y\{T^i_C,T^j_C\}=\tr T^i_C\{Y,T^j_C\}=\tr T^i_C\{T^j_C,Y\},
\nonumber\\&
\tr T^i_C\{T^j_C,T^k_C\}=A_{2C}\cdot d^{ijk}, \quad {\rm and}\quad \tr Y^3.
\label{TTT}
\end{eqnarray}
Expanding all possible cyclic permutation explicitly one reaches the following $\theta$-exact result for the SM gauge boson mixing coupling:
\begin{eqnarray}
\lefteqn{S_{\rm gauge}= S_{\rm gauge}^{\rm SM}}
\label{action3} \\
& &\hspace{-7mm}{}+{g^3_Y}\kappa_1\int {\theta^{\rho\tau}}\,
f^{\mu\nu}\left(\frac{1}{4}f_{\rho\tau}\star_2 f_{\mu\nu}-f_{\mu\rho}\star_2 f_{\nu\tau}\right),
 \nonumber \\
& &\hspace{-7mm}{}+g_Yg_L^2\kappa_2 \int\sum_{i=1}^{3} \theta^{\rho\tau}
\Big[B_i^{\mu\nu}\left(\frac{1}{4}f_{\rho\tau}\star_2B^i_{\mu\nu}-
2f_{\mu\rho}\star_2B^i_{\nu\tau}\right)
\nonumber \\
& &\hspace{2cm}{}
+f^{\mu\nu}\left(\frac{1}{2}B^i_{\rho\tau}\star_2B^{i}_{\mu\nu}-B^i_{\mu\rho}\star_2B^{i}_{\nu\tau}\right)
\Big],
 \nonumber \\
& &\hspace{-7mm}{}+g_Yg_C^2\kappa_3\int\sum_{j=1}^{8}
\theta^{\rho\tau}\Big[G_j^{\mu\nu}\left(\frac{1}{4}f_{\rho\tau}\star_2G^j_{\mu\nu}-2f_{\mu\rho}\star_2G^j_{\nu\tau}\right)
\nonumber \\
& &\hspace{2cm}{}
+f^{\mu\nu}\left(\frac{1}{2}G^j_{\rho\tau}\star_2G^j_{\mu\nu}-G^j_{\mu\rho}\star_2G^j_{\nu\tau}\right)
\Big] .
\nonumber
\end{eqnarray}
The couplings of the model $\kappa_i,\;\;i=1,2,3$, as functions of the six weights $\alpha_i $, are parameters of the model:
\begin{eqnarray}
\kappa_1&=&\frac{1}{2}\left(-\alpha_1-\frac{\alpha_2}{4}+\frac{8 \alpha_3}{9} -\frac{\alpha_4}{9}+\frac{\alpha_5}{36}+\frac{\alpha_6}{4}\right),
\nonumber\\
\kappa_2&=&\frac{1}{2}\Big(-\alpha_2+\alpha_5+\alpha_6\Big),
\nonumber\\
\kappa_3&=&\frac{1}{2} \Big(2\alpha_3-\alpha_4+\alpha_5\Big),
\label{kappa123}
\end{eqnarray}
with weights satisfying the positivity conditions,
\begin{equation} 
\alpha_j > 0,\;  \forall j,\;j=1,...,6.
\label{unequa}
\end{equation}
In order to restore the coupling constants, $\alpha_i$'s must satisfy three 
$\rm G_{SM}$ constraints \cite{Behr:2002wx}
\begin{eqnarray}
\frac{1}{g_Y^2}&=&2\alpha_1+\alpha_2+\frac{8\alpha_3}{3}+\frac{2\alpha_4}{3}+\frac{\alpha_5}{3}+\alpha_6,
\nonumber\\
\frac{1}{g_L^2}&=&\alpha_2+3\alpha_5+\alpha_6,
\nonumber\\
\frac{1}{g_C^2}&=&\alpha_3+\alpha_4+2\alpha_5.
\label{YLc123}
\end{eqnarray}
Solutions to the above system of six equations and six inequations are given in \cite{Behr:2002wx,Duplancic:2003hg}. Additional details of the model are given in \cite{Melic:2005fm,Buric:2006wm}. 

After performing the electroweak (EW) symmetry breaking, from the action (\ref{action3}) we extract  the $Z$ boson-photon and other TGB couplings in terms of the physical fields, which are not present in the commutative SM,
\begin{eqnarray}
{\cal L}_{Z\gamma\gamma}&=&\Big[g_Y^3\kappa_1\sin\vartheta_W\cos^2\vartheta_W
\nonumber\\
& &+g_Yg_L^2\kappa_2\big(\sin^3\vartheta_W
-2\sin\vartheta_W\cos^2\vartheta_W\big)\Big]\,
\nonumber\\
& & \cdot \;\theta^{\rho\tau} \Big[2Z^{\mu\nu}\big(2A_{\mu\rho}\star_2A_{\nu\tau}-A_{\mu\nu}\star_2A_{\rho\tau}\big)
\nonumber\\
& & +A^{\mu\nu}\big(8Z_{\mu\rho}\star_2A_{\nu\tau} -Z_{\rho\tau}\star_2A_{\mu\nu}\big)\Big],
 \label{L2}
\end{eqnarray}
where $A_{\mu\nu} = \partial_{\mu}A_{\nu}-\partial_{\nu}A_{\mu}$ and $Z_{\mu\nu} = \partial_{\mu}Z_{\nu}-\partial_{\nu}Z_{\mu}$, being the Abelian quantities.

Following the observations in \cite{Buric:2006wm,Buric:2007qx,Latas:2007eu,Horvat:2013rga}, one notice that introducing a further deformation in the gauge boson coupling may improve the quantum property of the model. This gauge deformation freedom $\kappa_g$ can be introduced into \eqref{L2} to reach a $\theta$-exact $Z\gamma\gamma$ coupling equivalent to the linear-in-$\theta$ one given in \cite{Buric:2006wm}, preserving, at the same time, the $\rm U(1)$ gauge symmetry. The modified interaction then read:
\begin{eqnarray}
{\cal L}_{Z\gamma\gamma}(\kappa_g)&=&\frac{e}{4} \sin2{\vartheta_W}\,{\rm K}_{Z\gamma \gamma}\,
{\theta^{\rho\tau}}
\label{L2mod}\\
&\cdot& \Big[2Z^{\mu\nu}\big(2A_{\mu\rho}\star_2A_{\nu\tau}-\kappa_gA_{\mu\nu}\star_2A_{\rho\tau}\big)
\nonumber\\
&&+  A^{\mu\nu}\big(8Z_{\mu\rho}\star_2A_{\nu\tau} -\kappa_gZ_{\rho\tau}\star_2A_{\mu\nu}\big)\Big],
\nonumber 
\end{eqnarray}
where the EW symmetry breaking, via some redefinitions, induces new dimensionless TGB coupling constant
\begin{equation}
K_{Z\gamma\gamma}=\frac{1}{2}\Big[g_Y^2\kappa_1+(g_Y^2-2g_L^2)\kappa_2\Big].
\label{KZgaga}
\end{equation}
Other interactions, like ${\cal L}_{\gamma\gamma\gamma}(\kappa_g)$, could be obtained in the same way.
The deformation parameter $\kappa_g$ is originally set to be one, however  $\kappa_g=3$ has been found to be improving the quantum property of several related models \cite{Buric:2006wm,Latas:2007eu,Horvat:2013rga}. Thus we shall do our phenomenological analysis of $Z\to\gamma\gamma$ decays for both values, i.e. for $\kappa_g=1,3$.
All new redefined dimensionless coupling constants ${\rm K}_{\gamma\gamma\gamma}$, ${\rm K}_{Z\gamma\gamma}$, ..., appears as a consequence of the EW symmetry breaking procedure, and were first introduced in \cite{Behr:2002wx,Duplancic:2003hg,Melic:2005fm}. It is important to note that
particularly interesting coupling ${\rm K}_{Z\gamma\gamma}$ could also
receive the zero value. See details in \cite{Behr:2002wx,Duplancic:2003hg}.

In this work we focus on the partial width of the
$Z\rightarrow \gamma\gamma$ decay arising from $\theta$-exact Lagrangian (\ref{L2mod}). The gauge-invariant amplitude ${\cal M}_{Z\rightarrow \gamma\gamma}$ for the $Z(k_1)\rightarrow\gamma(k_2)\,\gamma(k_3)$ decay
in the momentum space reads
\begin{eqnarray}
{\cal M}_{Z\to \gamma\gamma}&=&-2e \sin2{\vartheta_W}\,{\rm K}_{Z\gamma \gamma}F_{\star_2}(k_1,k_2)
\label{ampl}\\
&\cdot&{\Theta^{\mu\nu\rho}_3}(\kappa_g;k_1,-k_2,-k_3)
 \epsilon_{\mu}(k_1) \epsilon_{\nu}(k_2) \epsilon_{\rho}(k_3),
\nonumber
\end{eqnarray}
where momentum dependent function
\begin{eqnarray}
F_{\star_2}(k_1,k_2)=\frac{\sin\frac{k_1\theta k_2}{2}}{\frac{k_1\theta k_2}{2}},
\label{Ffactor}
\end{eqnarray}
comes from the $\star_2$-product in Lagrangian. The tensor ${\Theta^{\mu\nu\rho}_3}(\kappa_g;k_1,k_2,k_3)$, is given by
\begin{eqnarray}
{\Theta^{\mu\nu\rho}_3}(\kappa_g;k_1,k_2,k_3)&=&
-\,(k_1 \theta k_2)\,
\nonumber \\
& &
\hspace*{-3cm}
\cdot\;
\Big[(k_1-k_2)^\rho g^{\mu \nu} +(k_2-k_3)^\mu g^{\nu \rho} + (k_3-k_1)^\nu g^{\rho \mu}\Big]
\nonumber \\
& &
\hspace*{-3cm}
-\,\theta^{\mu \nu}\,
\Big[ k_1^\rho \, (k_2 k_3) - k_2^\rho \, (k_1 k_3) \Big]
\nonumber \\
& &
\hspace*{-3cm}
-\,\theta^{\nu \rho}\,
\Big[ k_2^\mu \, (k_3 k_1) - k_3^\mu \, (k_2 k_1) \Big]
\nonumber \\
& &
\hspace*{-3cm}
-\,\theta^{\rho \mu}\,
\Big[ k_3^\nu \, (k_1 k_2) - k_1^\nu \, (k_3 k_2) \Big]
\nonumber \\
& & \hspace*{-3cm}
+\,(\theta k_2)^\mu \,\Big[g^{\nu \rho}\, k_3^2 - k_3^\nu k_3^\rho\Big]
+(\theta k_3)^\mu\,\Big[g^{\nu \rho}\, k_2^2 - k_2^\nu k_2^\rho\Big]
\nonumber \\
& & \hspace*{-3cm}
+\,(\theta k_3)^\nu \,\Big[g^{\mu \rho}\, k_1^2 - k_1^\mu k_1^\rho \Big]
+(\theta k_1)^\nu \,\Big[g^{\mu \rho}\, k_3^2 - k_3^\mu k_3^\rho \Big]
\nonumber \\
& & \hspace*{-3cm}
+\,(\theta k_1)^\rho \,\Big[g^{\mu \nu}\, k_2^2 - k_2^\mu k_2^\nu \Big]
+(\theta k_2)^\rho \,\Big[g^{\mu \nu}\, k_1^2 - k_1^\mu k_1^\nu \Big]
\nonumber \\
& & \hspace*{-3cm}
+(\kappa_g-1)\,(\theta k_1)^{\mu}\,\Big[g^{\nu \rho}\,(k_3 k_2)-k_3^\nu k_2^\rho\Big]
\nonumber \\
& & \hspace*{-3cm}
+(\kappa_g-1)\,(\theta k_2)^{\nu} \,\Big[g^{\mu \rho}\,(k_3 k_1)-k_3^\mu k_1^\rho\Big]
\nonumber \\
& & \hspace*{-3cm}
+(\kappa_g-1)\,(\theta k_3)^{\rho} \,\Big[g^{\mu \nu}\,(k_2 k_1)-k_2^\mu k_1^\nu\Big]\, ,
\label{amplit}
\end{eqnarray}
where the 4-momenta $k_1,k_2,k_3$ are taken to be incoming, satisfying the momentum conservation $(k_1+k_2+k_3=0)$. In (\ref{amplit}) the deformation freedom parameter $\kappa_g$ appears symmetric in physical gauge bosons which enter the interaction point, as one would expect.  For $\kappa_g=1$, the tensor (\ref{amplit}) becomes  the tensor $\Theta_3((\mu,k_1),(\nu,k_2),(\rho,k_3))$ from \cite{Melic:2005fm}. It is straightforward to see that ${k_1}_\mu\Theta^{\mu\nu\rho}_3={k_2}_\nu\Theta^{\mu\nu\rho}_3={k_3}_\rho\Theta^{\mu\nu\rho}_3=0$, i.e. $\Theta^{\mu\nu\rho}_3(\kappa_g;k_1,k_2,k_3)$ respects the aforementioned $\rm U(1)$ gauge symmetry.

\section{Z Decays}

To illustrate certain physical effects of our deformed $\theta$-exact construction, we compute the $Z\to\gamma\gamma$ and $Z\to\nu\bar\nu$ decay rates in the Z--boson rest frame, which is then readily to be compared with the precision Z resonance measurements on $e^+e^-$ colliders, where Z is produced almost at rest.

\subsection{The NCSM $Z \to \gamma\gamma$ decay rate}
\noindent
The partial $Z \to \gamma\gamma$ decay width obtained from (\ref{ampl}) reads
\begin{equation}
\label{Z2Gamma:DecayWidthFull}
\begin{split}
&\Gamma(Z\to\gamma\gamma)=
\frac{ \alpha}{24} \sin^2 2\vartheta_W K_{Z\gamma\gamma}^2 M_Z
\\&
\cdot\;\Bigg[-
8\bigg(\Big(\kappa_g\big(9 \kappa_g-34\big) +35\Big) +2 \frac{|\vec{B_\theta}|^2}{|\vec{E_\theta}|^2}
\\&
+\big(\kappa_g-1\big) \big(\kappa_g+3\big) \frac{\big(\vec{E_\theta}\vec{B_\theta}\big)^2}{|\vec{E_\theta}|^4}\bigg)
\\&
+ \bigg(2 \Big(\kappa_g\big(11\kappa_g-42\big)+43\Big)
+\Big(\kappa_g\big(\kappa_g+2\big)+5\Big)\frac{|\vec{B_\theta}|^2}{ |\vec{E_\theta}|^2}
\\&
+\big(\kappa_g-1\big) \big(\kappa_g+3\big) \frac{\big(\vec{E_\theta}\vec{B_\theta}\big)^2}{|\vec{E_\theta}|^4}\bigg)
\\&
\cdot\,\bigg(M_Z^2|\vec{E_\theta}|
 \; {\rm Si}\Big(\frac{1}{2}M_Z^2|\vec{E_\theta}|\Big)
+2\cos\Big(\frac{1}{2}M_Z^2|\vec{E_\theta}|\Big)\bigg)
\\&
+ 2 \bigg(
-\big(\kappa_g-1\big)\big(\kappa_g+3\big)\frac{|\vec{B_\theta}|^2}{|\vec{E_\theta}|^2}
\Big(1-3\frac{\big(\vec{E_\theta}\vec{B_\theta}\big)^2}{|\vec{E_\theta}|^2|\vec{B_\theta}|^2}\Big)
\\&
+ 2\Big(\kappa_g\big(7\kappa_g-26\big)+27\Big)\bigg)
\frac{\sin\Big(\frac{1}{2}M_Z^2|\vec{E_\theta}|\Big)}{\Big(\frac{1}{2}M_Z^2|\vec{E_\theta}|\Big)}\Bigg],
\end{split}
\end{equation}
where we have used the following notation
$\theta^2 =(\theta^2)^{\mu}_{\mu} = \theta_{\mu\nu}\theta^{\nu\mu}
= 2\left (\vec{E}_{\theta}^2 - \vec{B}_{\theta}^2 \right )$, and
$|\vec{E_\theta}|\sim |\vec{B_\theta}|\sim1/\Lambda^2_{\rm NC}$.
In (\ref{Z2Gamma:DecayWidthFull}) Si is the sine integral function,
$\mbox{\rm Si}(z)= \int_0^z\,dt\, \frac{\sin t}{t}$. Expanding Si and trigonometric functions in power series, one can easily show that the partial width (\ref{Z2Gamma:DecayWidthFull}) recovers the prior leading order result in \cite{Buric:2007qx}
\begin{equation}
\begin{split}
\Gamma(Z\to\gamma\gamma)=&\frac{ \alpha}{72} \sin^2 2\vartheta_W K_{Z\gamma\gamma}^2 M_Z^5
\\&\cdot\Big[\big(13\kappa_g^2-50\kappa_g+51\big)|\vec{E_\theta}|^2
\\&+\big(\kappa_g^2+2\kappa_g+3\big)|\vec{B_\theta}|^2\Big]+\mathcal{O}\left(\Lambda_{\rm NC}^{-8}\right)
\label{Z2GammaDecay:a}
\end{split}
\end{equation}

The $\theta$-exact rate (\ref{Z2Gamma:DecayWidthFull}) differs considerably from the rate (\ref{Z2GammaDecay:a}) obtained at the $\theta$-first order in the SW/enveloping algebra $\theta$-expanded model \cite{Buric:2007qx}. Difference appears due to the simultaneous presence of  the $\theta$-exact model function $F_{\star_2}(k_1,k_2)$, and of the deformation freedom parameter $\kappa_g$. Both objects ($F_{\star_2}$ and $\kappa_g$) together produce new, previously unknown, mixing term:
$(\kappa_g-1) (\kappa_g+3)(\vec{E_\theta}\vec{B_\theta})^2$, which vanishes for $\kappa_g=1,-3$, $\vec{E_\theta}=0$, $\vec{B_\theta}=0$ or $\vec{E_\theta}\perp \vec{B_\theta}$. So, unlike the first order rate which depends only on the lengths of $\vec{E}_\theta$ and $\vec{B}_\theta$, the $\theta$-exact rate has one more higher order rotation invariant term $(\vec E_\theta\vec B_\theta)^2$, which would measure the relative angle between $\vec{E}_\theta$ and $\vec{B}_\theta$. At the traditional (undeformed gauge interaction) point $\kappa_g=1$ one would not see this effect.

In the $Z$-rest frame the $Z$-momenta are reduced to the time-like component only, therefore $k_1\theta k_2=M_Z {\vec E}_\theta \cdot \vec k_2$. Thus the ${\vec B}_\theta$ component does not contribute to the $F_{\star_2}$-function. Consequently, the $\theta$-exact rate equals to the first order result for space-like noncommutativity as expected. For light-like noncommutativity (also preserving unitarity \cite{Aharony:2000gz}) the full NC effect will be still exhibited.

\subsection{The NCSM $Z\to \nu\bar{\nu}$ decay rate }
\noindent
Since the complete, $\theta$-exact $Z\nu\nu$ interaction on noncommutative spaces was discussed in details in \cite{Horvat:2011qn}, we shall not repeat it here. We only give the $Z\nu\bar\nu$ vertex, which was first time derived in our previous paper \cite{Horvat:2011qn}:
\begin{equation}
\label{vertexznunubar}
\begin{split}
\Gamma&_{Z\nu^l\bar\nu^{l'}}=
i\frac{e}{\sin 2\vartheta_W}\Bigg\{\bigg[\gamma^\mu
+\frac{i}{2}F_{\bullet}(q,k_l)
\\&\cdot \sum\limits_{n=1}^{3}U^*_{l'n}U_{ln}\bigg((q\theta k_l)\gamma^\mu
+(\theta q)^\mu\fmslash k_l-(\theta k_l)^{\mu}\fmslash q\bigg)\bigg]\frac{1-\gamma^5}{2}
\\&
+\sum\limits^3_{n=1}\sum\limits^{N+3}_{n'=4}
\left((\theta k_l)^{\mu}U_{l'n}(m^D_{n(n'-3)})^*U_{ln'}\frac{1-\gamma^5}{2}
\right.\\
&\left.
+\Big((\theta q)^\mu-(\theta k_l)^\mu\Big) U^*_{l'n}m^D_{n(n'-3)}U^*_{ln'}
\frac{1+\gamma^5}{2}\right)\bigg\}\\
&+\frac{\kappa e}{2}\tan\vartheta_W \bigg\{F_{\star_2}(q,k_l)\delta_{ll'}\bigg[(q\theta k_l)\gamma^\mu
\\&
+(\theta q)^\mu(\fmslash k_l-m_l)-(\theta k_l)^\mu\fmslash q\bigg]\bigg\}\,,
\end{split}
\end{equation}
where $U$ is the mixing matrix, and $m^D$ denotes the Dirac mass part of the mass matrix
\begin{equation}
{\bf M}=\begin{pmatrix}
0 & m^D\\
(m^D)^T & m^M
\end{pmatrix}\,.
\label{mmatrixseesaw1}
\end{equation}
Indices $l,l'$ runs from $1$ to $N+3$ and denote the $N+3$ neutrino mass eigenstates.
The momentum dependent factor
\begin{equation}
F_{\bullet}(q,k_l):=\frac{(e^{-i\frac{q\theta k_l}{2}}-1)}{-i\frac{q\theta k_l}{2}}.
\label{factorbullet}
\end{equation}
arising from generalized $\star$-products, while the constant $\kappa$ measures a correction from the chiral blind $\star$-commutator coupling between neutrinos and $Z$ boson. It vanishes in the commutative limit therefore may be arbitrarily large. The non-$\kappa$-proportional term, on the other hand, is the noncommutative deformation of standard model Z-neutrino coupling, which involves the left handed neutrinos only. For details see  section four in \cite{Horvat:2011qn}.

For simplicity we set all neutrino masses to be zero in this paper. Then for massless on-shell neutrinos
the terms
\begin{equation}
[ (\theta q)^{\mu} {\fmslash k_l}
-(\theta k_l)^\mu {\fmslash q} ]\, (1-\gamma_5),
\label{p'p}
\end{equation}
in the vertex (\ref{vertexznunubar}) do not contribute to the $Z\to\nu\bar\nu$ amplitude due to the equations of motions. Thus the vertex (\ref{vertexznunubar}) can be written in a form similar to the SM vertex
\begin{equation}
\frac{ig}{2\cos\vartheta_W}\,\gamma^\mu\,(g_V-g_A\gamma_5),
\label{gVgA}
\end{equation}
 with
\begin{eqnarray}
g_V &=& 1
   -\frac{1}{2}
   \exp\Big(i\frac{M_Z}{2} \, \vec p\vec E_\theta \Big)
   \nonumber\\
  & +&
2i\kappa \sin^2\vartheta_W  \sin\Big(\frac{M_Z}{2} \, \vec p\vec E_\theta \Big)~,
\label{Z2fbarfNC:gV} \\
g_A &=& 1 -\frac{1}{2}
\exp\Big(i\frac{M_Z}{2} \, \vec p\vec E_\theta \Big)~.
\label{Z2fbarfNC:gA}
\end{eqnarray}

Using above vertex we obtain the following $Z\to \nu\bar{\nu}$ decay partial width
\begin{equation}
\begin{split}
&\Gamma(Z\to\nu\bar\nu)=
\Gamma_{\rm SM}(Z\to\nu\bar\nu)+\Gamma_{\kappa}(Z\to\nu\bar\nu)
\\&
=\Gamma_{\rm SM}(Z\to\nu\bar\nu)
\\&
+\frac{\alpha}{3 M_Z |\vec{E_\theta}|}
   \bigg[\kappa  \big(1 -\kappa +\kappa\cos 2 \vartheta _W \big)
   \sec ^2\vartheta_W\cos\Big(\frac{1}{4}M_Z^2 |\vec{E_\theta}|\Big)
\\&
   -8 \csc ^2 2 \vartheta _W\bigg]\sin\Big(\frac{1}{4}M_Z^2 |\vec{E_\theta}|\Big)
\\&
+\frac{\alpha M_Z}{12}   \bigg[-2 \kappa ^2
+\big(\kappa  (2 \kappa -1)+2\big) \sec^2\vartheta _W+2 \csc^2\vartheta _W\bigg]
\\&=\Gamma_{\rm SM}(Z\to\nu\bar\nu)+\frac{\alpha M_Z^5|\vec{E_\theta}|^2}{288}\bigg[2 \csc ^2 2 \vartheta _W
\\&+\kappa\big((2\kappa-1)\sec ^2\vartheta_W -2\kappa\big)\bigg]+\mathcal{O}\left(\Lambda_{\rm NC}^{-8}\right),
\end{split}
\label{rateZnunu}
\end{equation}
whose NC part vanishes when $\vec{E}_\theta\to 0$, i.e. for vanishing $\theta$ or space-like noncommutativity, but not light-like. As before, in the $Z$-rest frame the $Z$-momenta are reduced to the time-like component only, therefore $q\theta k=M_Z {\vec E}_\theta \vec k$, and the ${\vec B}_\theta$ component does not contribute to the $F_{\star_2}$ and $F_{\bullet}$ functions, respectively.


\section{Discussion and conclusion}

In the last section we have shown that the tree level tri-particle decays ($Z\to\gamma\gamma,\;\nu\bar\nu$) in the covariant noncommutative quantum gauge theory based on Seiberg-Witten maps can be computed without an expansion over the noncommutative parameter $\theta$. We obtained for the first time covariant $\theta$-exact triple neutral gauge boson interactions within the NCSM gauge sector. For computation of the invisible $Z$-decays we have also needed to use $\theta$-exact $Z\nu\bar\nu$ interactions constructed for the first time in \cite{Horvat:2011qn}.

Focusing on Z decays into two photons and two neutrinos, we have reconsidered previous computations that were done with less sophisticated tools and derived new bounds on the scale of noncommutativity. Let us present here our phenomenological results and compare to those obtained previously.  Noncommutative, especially the $\theta$-exact computations are always prone to the question of nonlocal quantum corrections. However several prior results~\cite{Buric:2006wm,Latas:2007eu,Buric:2007ix,Martin:2009vg,Martin:2009sg, Tamarit:2009iy,Buric:2010wd,Horvat:2011bs,Horvat:2011qg,Horvat:2013rga} have shown that the severity of the novel divergent behavior can be put under control by adjusting the aforementioned deformation parameter(s). We are positive that such a control could be 
extended across an even broader spectrum. And the tree level amplitude evaluation is of considerable phenomenological importance all the time. In the following we include certain variations of the deformation parameter to illustrate its phenomenological effect too.

(I) First, we shall perform our analysis of $Z\to\gamma\gamma$ decays for the two unitarity preserving cases: that is for the light-like and for the space-like noncommutativity, respectively. Second, since the divergent quantum properties of our NCSM gauge sector are controlled by two deformation parameter highlighted values: $\kappa_g=1,3$, we shall do our numerical analysis for those two values, too.

Now we define the ratio
$\Gamma(Z\to\gamma\gamma)/\Gamma(Z)_{\rm{tot,SM}}$ where $\Gamma(Z)_{\rm{tot,SM}}=(2.4952 \pm  0.0023)$ GeV is the Z--boson full width \cite{PDG2011}. The Z--boson mass is $M_Z=(91.1876\pm 0.0021)$ GeV, the Weinberg angle $\sin^2\vartheta_W = 0.23116$, and the fine structure constant $\alpha$ equals $1/137.036$, \cite{PDG2011}.

\begin{figure}[top]
\begin{center}
\includegraphics[width=8.5cm,angle=0]{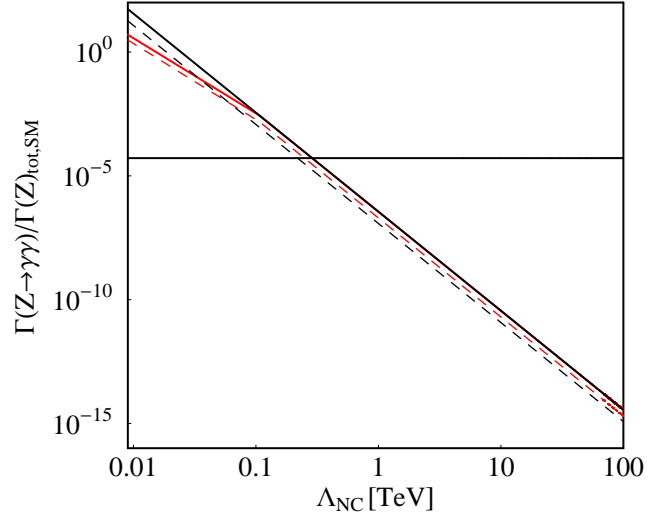}
\end{center}
\caption{$\Gamma(Z\to\gamma\gamma)/\Gamma(Z)_{\rm{tot,SM}}$ vs. $\Lambda_{\rm NC}$, for fixed coupling constant $|K_{Z\gamma\gamma}|=0.33$.
The black horizontal line is the experimental upper limit $\Gamma(Z\to\gamma\gamma)/\Gamma(Z)_{\rm{tot,SM}}<5.2\cdot 10^{-5}$ [64]. Dashed and solid curves correspond to the gauge deformation freedom parameter $\kappa_g=1,3$, respectively.
Red corresponds to the light-like case $|\vec{E_\theta}|=|\vec{B_\theta}|=1/\sqrt{2}\Lambda_{\rm{NC}}^2$ and $\vec{E_\theta}\vec{B_\theta}=0$,
(overlapped with $\vec{E_\theta}\vec{B_\theta}=1/2\Lambda_{\rm{NC}}^4$). Black is:  $|\vec{E_\theta}|=\vec{E_\theta}\vec{B_\theta}=0$, and $|\vec{B_\theta}|=1/\Lambda_{\rm{NC}}^2$.}
\label{Fig1}
\end{figure}

Figure \ref{Fig1} displays main results for the ratio $\Gamma(Z\to\gamma\gamma)/\Gamma(Z)_{\rm{tot,SM}}$ based on Eq.~(\ref{Z2Gamma:DecayWidthFull}), as a function of the scale of noncommutativity $\Lambda_{\rm NC}$, for fixed coupling constant $|K_{Z\gamma\gamma}|=0.33$, and for two choices of deformation parameter $\kappa_g=1,3$, respectively.

\begin{figure}[top]
\begin{center}
\includegraphics[width=8.5cm,angle=0]{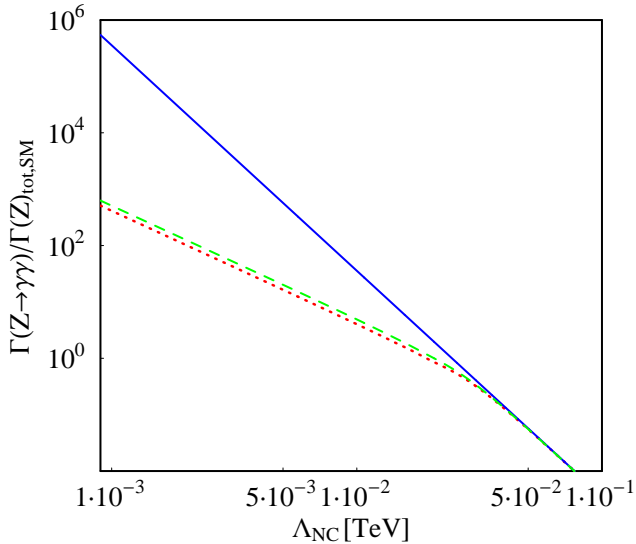}
\end{center}
\caption{$\Gamma(Z\to\gamma\gamma)/\Gamma(Z)_{\rm{tot,SM}}$ vs. $\Lambda_{\rm NC}$ at very small $\Lambda_{\rm NC}$ scales, and for fixed coupling $|K_{Z\gamma\gamma}|=0.33$, and freedom $\kappa_g=3$. Here the space-like case
$|\vec{E_\theta}|=\vec{E_\theta}\vec{B_\theta}=0$, $|\vec{B_\theta}|=1/\Lambda_{\rm{NC}}^2$, and which equals to the first order approximation for space-like, light-like and full rank $\theta$ is given in solid (blue) line. The light-like case, $|\vec{E_\theta}|=|\vec{B_\theta}|=1/\sqrt{2}\Lambda_{\rm{NC}}^2,\;\vec{E_\theta}\perp \vec{B_\theta}$ is given in dotted (red) and full rank $\theta$, $|\vec{E_\theta}|=|\vec{B_\theta}|=1/\sqrt{2}\Lambda_{\rm{NC}}^2,\;\vec{E_\theta}\parallel \vec{B_\theta}$ in dashed (green) curve, respectively. Small but visible deviation can be seen between dotted and dashed curves.}
\label{Fig2}
\end{figure}

Figure ~\ref{Fig2} magnifies the low NC scale regime of Fig ~\ref{Fig1} for fixed coupling constant $|K_{Z\gamma\gamma}|=0.33$, for fixed gauge-deformation parameter $\kappa_g=3$, and with one more curve for the full rank $\theta^{\mu\nu}$ \cite{Horvat:2013rga} added.\footnote{The unitarity issue in the one-loop photon two point function of \cite{Gomis:2000zz,Aharony:2000gz}, is shown to be absent for the same full rank $\theta$ in the U(1) model~of \cite{Horvat:2013rga} with deformation $\kappa_g=3$.} Note that the first order result for $\kappa_g=3$ becomes
\begin{equation}
\begin{split}
\Gamma(Z\to\gamma\gamma)=&\frac{ \alpha}{4} \sin^2 2\vartheta_W K_{Z\gamma\gamma}^2 M_Z^5\Big(|\vec{E}_\theta|^2+|\vec{B}_\theta|^2\Big)
\\&+\mathcal{O}\big(\Lambda_{\rm NC}^{-8}\big),
\end{split}
\end{equation}
with the two lengths $|\vec{E}_\theta|$ and $|\vec{B}_\theta|$ weighing exactly the same. Therefore the first order result for all three $\theta$ choices in Figure ~\ref{Fig2} coincide with the full result for space-like $\theta$, while the $\theta$-exact results access the difference among $\theta^{\mu\nu}$ choices at low NC scales.

An important consequence of this paper is that the coupling $K_{Z\gamma\gamma}$ and the NC scale $\Lambda_{\rm NC}$ are independent parameters, whereas in the $\theta$-expanded  models (linear in $\theta$) experiments measure always the combination $|K_{Z\gamma\gamma}|^2/\Lambda_{\rm NC}^4$.


Furthermore, at very $\Lambda_{\rm NC}$ low scales the full $\theta$-exact decay rate is sensitive to the tensor structure of the $\theta^{\mu\nu}$ tensor for $\kappa_g\neq 1, -3$, as shown in Fig~\ref{Fig2}. However, these differences are not relevant when the NC scale is much larger than the $Z$-boson mass. This fact is quite transparent from both Figs, \ref{Fig1}, and ~\ref{Fig2}, respectively. In order to maximize the $Z\to\gamma\gamma$ rate, in the above analyzes we have used the maximal allowed value $|K_{Z\gamma\gamma}|=0.33$ computed and given in figures and tables of \cite{Duplancic:2003hg}. This is different with respect to the analysis of the rate computed at the first order in $\theta$ in~\cite{Buric:2007qx}, where the lower central value from the figures and the tables in \cite{Duplancic:2003hg} $|K_{Z\gamma\gamma}|=0.05$ was used. Anyhow the ratio of two values $|0.33/0.05|^2\simeq43$, is representing just an overall shift-factor, in any analysis. Finally it is important to note that coupling $K_{Z\gamma\gamma}$ could also receive the zero value, thus producing zero rate for the $Z \to\gamma\gamma$ decay.
\\
\\

Inspecting Fig.~\ref{Fig1} we see that current experimental upper limit $\Gamma(Z\to\gamma\gamma)/\Gamma(Z)_{\rm{tot,SM}}<5.2\cdot 10^{-5}$ \cite{PDG2011} is way too weak to produce any meaningful constraint on the scale of noncommutativity.
However, we can certainly expect that further analysis of the LHC experiments would significantly improve the current limit on the $Z\to\gamma\gamma$ partial width. Also Fig. \ref{Fig1} clearly shows that, for example, for ratio as low as $\sim10^{-15}$, our noncommutative scale is $\Lambda_{\rm{NC}} \stackrel{>}\sim 100$ TeV,  thus unobservable at the LHC energies.

\begin{figure}[top]
\begin{center}
\includegraphics[width=8.5cm,angle=0]{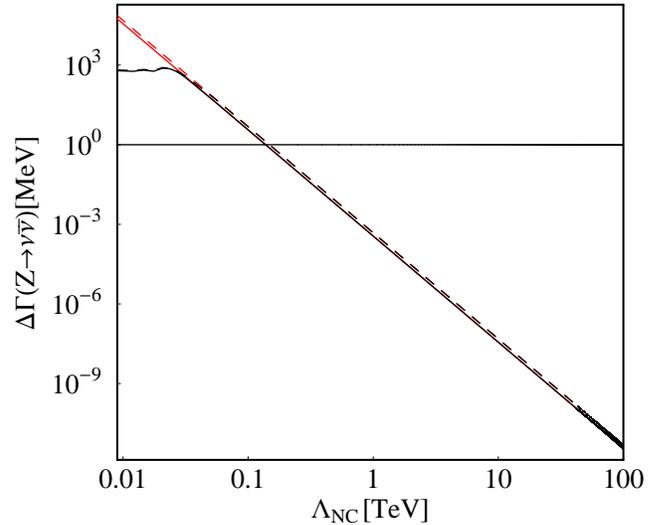}
\end{center}
\caption{$\Delta\Gamma(Z\to\nu\bar\nu)$ decay width vs. $\Lambda_{\rm{NC}}$, for $\kappa=1$ (solid) and $\kappa=0$ (dashed), respectively. Black curves correspond to $\theta$-exact rate, while straight (red) curves correspond to the expanded decay rate.}
\label{Fig3}
\end{figure}

(II) Solely new analysis of the NC contributions to the $Z\to\nu\bar\nu$ decay mode rate (\ref{rateZnunu}) is presented next. A comparison of the experimental $Z\to\nu\bar\nu$ decay width $\Gamma_{\rm invisible}=(499.0\pm 1.5)$ MeV \cite{PDG2011} with its SM theoretical counterpart, allows us to set a constraint $\Delta\Gamma=\Gamma(Z\to\nu\bar\nu) - \Gamma_{\rm SM}(Z\to\nu\bar\nu) \lesssim 1$ MeV, from where a bound on the scale of noncommutativity
$\Lambda_{\rm{NC}} = {|\vec{E_\theta}|^{-1/2}}\stackrel{>}\sim 120$ GeV is obtained  (see Fig.~\ref{Fig3}), for both choices $\kappa =0,1$.

It is worth to notice that since (\ref{rateZnunu}) depends only on $E_{\theta}$ the effect exists only for $E_{\theta}\neq 0$. Thus Fig. \ref{Fig3} does not include the space-like noncommutativity case.

Comparing to previous results, the total decay rates are modified by a factor which remains finite throughout all energy scales. Thus, our results behave much better than the $\theta$-expansion method when ultra high energy processes are considered. We expect that similar control on the high energy behavior can be extended to $\theta$-exact perturbation theory involving more external fields in the near future. All of our results show closed/convergent forms (see Figs. ~\ref{Fig1} to \ref{Fig3}) throughout full interaction energy scales, thus facilitating further phenomenological applications. This would provide a considerably improved theoretical basis for research work in the field of noncommutative particle phenomenology. \\

The work of R.H., D.K., J.T., and J.Y. are supported by the Croatian Ministry of Science,
Education and Sports under Contract No. 098-0982930-2872.
The work of A.I. supported by the Croatian Ministry of Science, Education and Sports under Contracts No.
098-0982930-1016.


\begin{thebibliography}{99}

\bibitem{Seiberg:1999vs}
N.~Seiberg and E.~Witten, 
  JHEP  {\bf 9909} (1999) 032.

\bibitem{Hinchliffe:2002km}
I.~Hinchliffe, N.~Kersting, and Y.~L. Ma,
  {\em Int. J. Mod. Phys.}  {\bf A19} (2004).

\bibitem{Trampetic:2008bk}
  J.~Trampetic,
  Fortsch.\ Phys.\  {\bf 56} (2008) 521.


\bibitem{Jackiw:2001jb}
  R.~Jackiw, S.~Y.~Pi,
  {\em Phys.\ Rev.\ Lett.}  {\bf 88} (2002) 111603.

\bibitem{Madore:2000en}
J.~Madore et al,
  {\em Eur. Phys. J.} {\bf C16} (2000) 161.

\bibitem{Jurco:2000ja}
B.~Jurco et al,
{\em Eur. Phys. J.} {\bf C17} (2000) 521.

\bibitem{Jurco:2000fb}
  B.~Jurco and P.~Schupp,
  {\em Eur.\ Phys.\ J.}  {\bf C14} (2000) 367.

\bibitem{Jurco:2001my}
B.~Jurco, P.~Schupp, and J.~Wess, 
  {\em Nucl. Phys.} {\bf B604} (2001) 148.

\bibitem{Jurco:2001rq}
B.~Jurco et al
  {\em Eur. Phys. J.} {\bf C21} (2001) 383.

\bibitem{Calmet:2001na}
X.~Calmet et al
   {\em Eur. Phys. J.} {\bf C23} (2002) 363.

\bibitem{Deshpande:2001mu}
  N.~Deshpande and X.~He,
  {\em Phys. Lett.} {\bf B533} (2002) 116.

\bibitem{Horvat:2011qn}
R.~Horvat et al,
{\em Phys. Lett.} {\bf B715}, 340-347 (2012).

\bibitem{Chaichian:2009uw}
 M.~Chaichian et al,
 {\em  Phys. Lett.} {\bf B683}, 55-61 (2010).

\bibitem{Behr:2002wx}
W.~Behr et al,
  {\em Eur. Phys. J.} {\bf C29} (2003) 441.
%
\bibitem{Duplancic:2003hg}
G.~Duplan\v{c}i\'{c}, P.~Schupp and J.~Trampeti\'{c},
Eur.~Phys. J. C{\bf 32} (2003) 141.

\bibitem{Aschieri:2002mc}
P.~Aschieri et al,
{\em Nucl. Phys.} {\bf B651} (2003) 45.

\bibitem{Melic:2005fm}
B.~Melic et al,
  {\em Eur. Phys. J.} {\bf C42} (2005) 483, ibid 499.

\bibitem{Martin:2013gma}
  C.~P.~Martin,
  Class.\ Quant.\ Grav.\  {\bf 30} (2013) 155019.

\bibitem{Martin:2013lba}
  C.~P.~Martin,
  arXiv:1311.2826 [hep-th].


\bibitem{Bichl:2001cq}
A.~Bichl et al,
  JHEP {\bf 0106} (2001) 013.

\bibitem{Ohl:2004tn}
T.~Ohl and J.~Reuter,
  {\em Phys. Rev.} {\bf D70} (2004) 076007.

\bibitem{Melic:2005su}
B.~Melic, K.~Passek-Kumericki, and J.~Trampetic,
  {\em Phys. Rev.} {\bf D72} (2005) 057502.

\bibitem{Alboteanu:2006hh}
A.~Alboteanu, T.~Ohl, and R.~Ruckl,
{\em Phys. Rev.} {\bf D74} (2006) 096004, ibid {\em Acta Phys. Polon.} {\bf B38} (2007) 3647.



\bibitem{Buric:2007qx}
M.~Buric et al,
  {\em Phys. Rev.} {\bf D75} (2007) 097701.

\bibitem{Filk:1996dm}
T.~Filk, 
{\em Phys. Lett.} {\bf B376} (1996) 53.


\bibitem{Grosse:2004yu}
H.~Grosse and R.~Wulkenhaar, 
  Commun. Math.Phys. {\bf 256} (2005) 305.

\bibitem{Magnen:2008pd}
  J.~Magnen, V.~Rivasseau and A.~Tanasa,
  Europhys.\ Lett.\  {\bf 86} (2009) 11001.
  %
\bibitem{Meljanac:2011cs}
  S.~Meljanac et al,
  JHEP {\bf 1112} (2011) 010.

\bibitem{Vilar:2009er}
L.~C.~Q. Vilar et al,
{\bf PoS ISFTG (2009) 071}.


\bibitem{Martin:2002nr}
C.~P. Martin, 
{\em Nucl. Phys.} {\bf B652} (2003) 72.

\bibitem{Buric:2006wm}
M.~Buric, V.~Radovanovic, and J.~Trampetic,
  {JHEP} {\bf  0703} (2007) 030.

\bibitem{Latas:2007eu}
D.~Latas, V.~Radovanovic, and J.~Trampetic,
  {\em Phys. Rev.} {\bf D76} (2007) 085006.

\bibitem{Buric:2007ix}
  M.~Buric et al,
 {\em  Phys. Rev.} {\bf D77} (2008) 045031.
  %
\bibitem{Martin:2009sg}
  C. P. Martin, C.~Tamarit,
 {\em  Phys. Rev.} {\bf D80}, 065023 (2009).
  %
\bibitem{Martin:2009vg}
  C.~P.~Martin, C.~Tamarit,
  JHEP {\bf 0912} (2009) 042.
    %
\bibitem{Tamarit:2009iy}
  C.~Tamarit,
  {\em Phys. Rev.} {\bf D81} (2010) 025006.
  %
\bibitem{Buric:2010wd}
  M.~Buric et al,
  {\em Phys. Rev.} {\bf D83} (2011) 045023.

\bibitem{Schupp:2002up}
P.~Schupp et al,
   {\em Eur. Phys. J.} {\bf C36} (2004) 405.

\bibitem{Minkowski:2003jg}
P.~Minkowski, P.~Schupp, and J.~Trampetic,
  {\em Eur. Phys. J.} {\bf C37}  (2004) 123.

\bibitem{Horvat:2009cm}
  R.~Horvat, J.~Trampetic,
  {\em Phys. Rev.} {\bf D79} (2009) 087701.

\bibitem{Horvat:2010sr}
  R.~Horvat, D.~Kekez and J.~Trampetic,
 {\em  Phys. Rev.} {\bf D83} (2011) 065013.
  %
\bibitem{Horvat:2011iv}
  R.~Horvat et al,
{\em Phys. Rev.} {\bf D84} (2011) 045004.

\bibitem{Horvat:2011wh}
  R.~Horvat, J.~Trampetic,
{\em Phys. Lett.} {\bf B710} (2012) 219.

\bibitem{Mehen:2000vs}
T.~Mehen and M.~B. Wise,
JHEP {\bf 0012} (2000) 008.

\bibitem{Liu:2000mja}
H.~Liu,
  {\em Nucl. Phys.} {\bf B614} (2001) 305.

\bibitem{Okawa:2001mv}
Y.~Okawa and H.~Ooguri,
  {\em Phys. Rev.} {\bf D64} (2001) 046009.

\bibitem{Martin:2012aw}
  C.~P.~Martin,
{\em Phys. Rev.} {\bf D86} (2012) 065010.

\bibitem{Schupp:2008fs}
P.~Schupp and J.~You,
 JHEP {\bf 0808} (2008) 107.

\bibitem{Horvat:2011bs}
  R.~Horvat et al,
JHEP {\bf 1112} (2011) 081.

\bibitem{Horvat:2011qg}
  R.~Horvat et al,
 JHEP {\bf 1204} (2012) 108.
  %
\bibitem{Horvat:2010km}
  R.~Horvat and J.~Trampetic,
  JHEP {\bf 1101} (2011) 112.

\bibitem{Horvat:2013rga}
  R.~Horvat et al, 
   JHEP {\bf 1311} (2013) 071.

\bibitem{hep-th/0606248}
  S.~Abel, C.~-S.~Chu and M.~Goodsell,
JHEP\ {\bf 0611} (2006) 058. 
  %
\bibitem{Abel:2006wj}
  S.~A.~Abel et all, 
  JHEP {\bf 0609}, 074 (2006).
  %
\bibitem{LandauYoung}
L.D. Landau, Dokl. Akad. Nauk. USSR 60, 242 (1948);
C.N. Yang, {\em Phys. Rev.} {\bf 77}, 242 (1950).
  %
\bibitem{SheikhJabbari:2000vi}
  M.~M.~Sheikh-Jabbari,
  {\em Phys. Rev. Lett.} {\bf 84}, 5265 (2000).
%
\bibitem{Melic:2005hb}
  B.~Melic, K.~Passek-Kumericki and J.~Trampetic,
 {\em  Phys. Rev.} {\bf D72}, 054004 (2005).
\bibitem{Tamarit:2008vy}
  C.~Tamarit and J.~Trampetic,
  {\em Phys. Rev.} {\bf D79}, 025020 (2009).

\bibitem{Gomis:2000zz}
  J.~Gomis and T.~Mehen,
 {\em  Nucl.\ Phys.} {\bf B591}, 265 (2000).

\bibitem{Aharony:2000gz}
  O.~Aharony, J.~Gomis and T.~Mehen,
  JHEP {\bf 0009}, 023 (2000).


\bibitem{Pati:1974yy}
  J.~C.~Pati and A.~Salam,
 {\em  Phys. Rev.}  {\bf D10} (1974) 275
   [Erratum-ibid.\  {\bf D11} (1975) 703].

\bibitem{trinification}
  S.~L.~Glashow,
 {\it Proc. Fourth Workshop(1984)
on Grand Unification}, ed. K. Kang {\it et. al.}
(World Scientific, Singapore, 1985), p.88.

\bibitem{Dias:2010vt}
  A.~G.~Dias et al,
   {\em Phys. Rev.}  {\bf D82} (2010) 035013.

   %
\bibitem{PDG2011}
J. Beringer et al. (Particle Data Group), Phys. Rev. {\bf D86}, 010001 (2012),
  and 2013 partial update for the 2014 edition.




%

\end{thebibliography}
\end{document}